\documentclass{PoS}

\usepackage[utf8]{inputenc}
\usepackage{wrapfig}

\title{Features of W production in p-p, p-Pb and Pb-Pb collisions}

\ShortTitle{Features of W production in p-p, p-Pb and Pb-Pb collisions}

\author{Fran\c{c}ois Arleo \\
Laboratoire Leprince-Ringuet, \'Ecole polytechnique, CNRS/IN2P3 91128 Palaiseau, France. \\
E-mail: \email{francois.arleo@cern.ch}}

\author{\'Emilien Chapon \\
CERN, Experimental Physics Department, CERN, CH-1211 Geneva 23, Switzerland, \\
Laboratoire Leprince-Ringuet, \'Ecole polytechnique, CNRS/IN2P3 91128 Palaiseau, France. \\
E-mail: \email{emilien.chapon@cern.ch}}

\author{\speaker{Hannu Paukkunen} \\
University of Jyvaskyla, Department of Physics, P.O. Box 35, FI-40014 University of Jyvaskyla, Finland \\
Helsinki Institute of Physics, P.O. Box 64, FI-00014 University of Helsinki, Finland \\
Instituto Galego de F\'\i sica de Altas Enerx\'\i as (IGFAE), Universidade de Santiago de Compostela, E-15782 Galicia, Spain. \\
E-mail: \email{hannu.paukkunen@jyu.fi}}

\abstract{We consider the production of inclusive W bosons in variety of high-energy hadronic collisions: p--p, p--$\overline{\rm p}$, p--Pb, and Pb--Pb. In particular, we focus on the resulting distributions of charged leptons from W decay that can be measured with relatively low backgrounds. The leading-order expressions within the collinearly factorized QCD indicate that the center-of-mass energy dependence at forward/backward rapidities should be well approximated by a simple power law. The scaling exponent is related to the small-$x$ behaviour of the quark distributions, which is largely driven by the parton evolution. An interesting consequence is the simple scaling law for the lepton charge asymmetry which relates measurements in different collision systems. The expectations are contrasted with the existing data and a very good overall agreement is found. Finally, we propose precision observables to be measured at the LHC.}

\FullConference{XXV International Workshop on Deep-Inelastic Scattering and Related Subjects\\
		3-7 April 2017\\
		University of Birmingham, UK}

\begin{document}

\section{Theoretical background}

\begin{figure}[htb!]
\center
\includegraphics[width=0.45\textwidth]{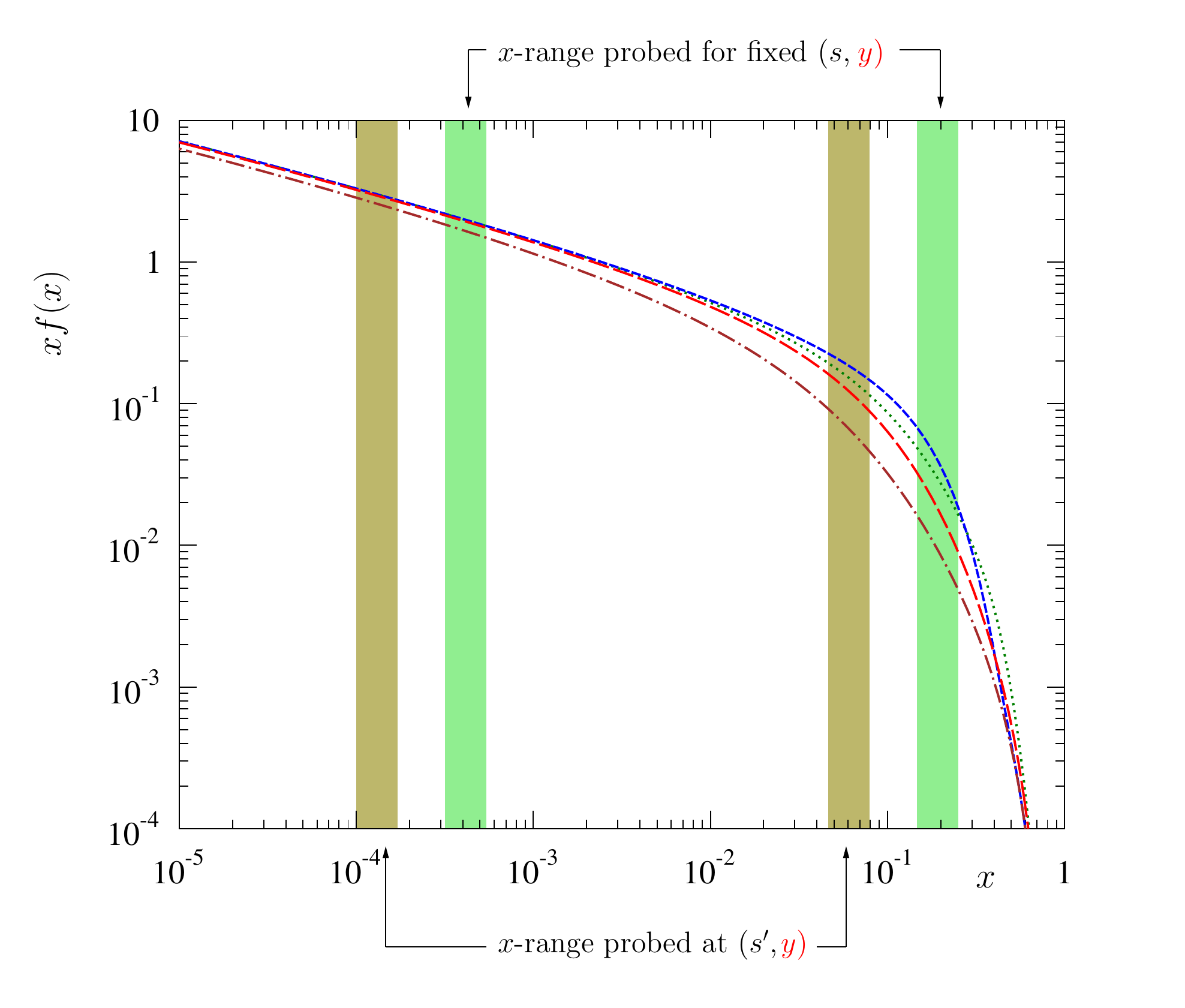}
\includegraphics[width=0.45\textwidth]{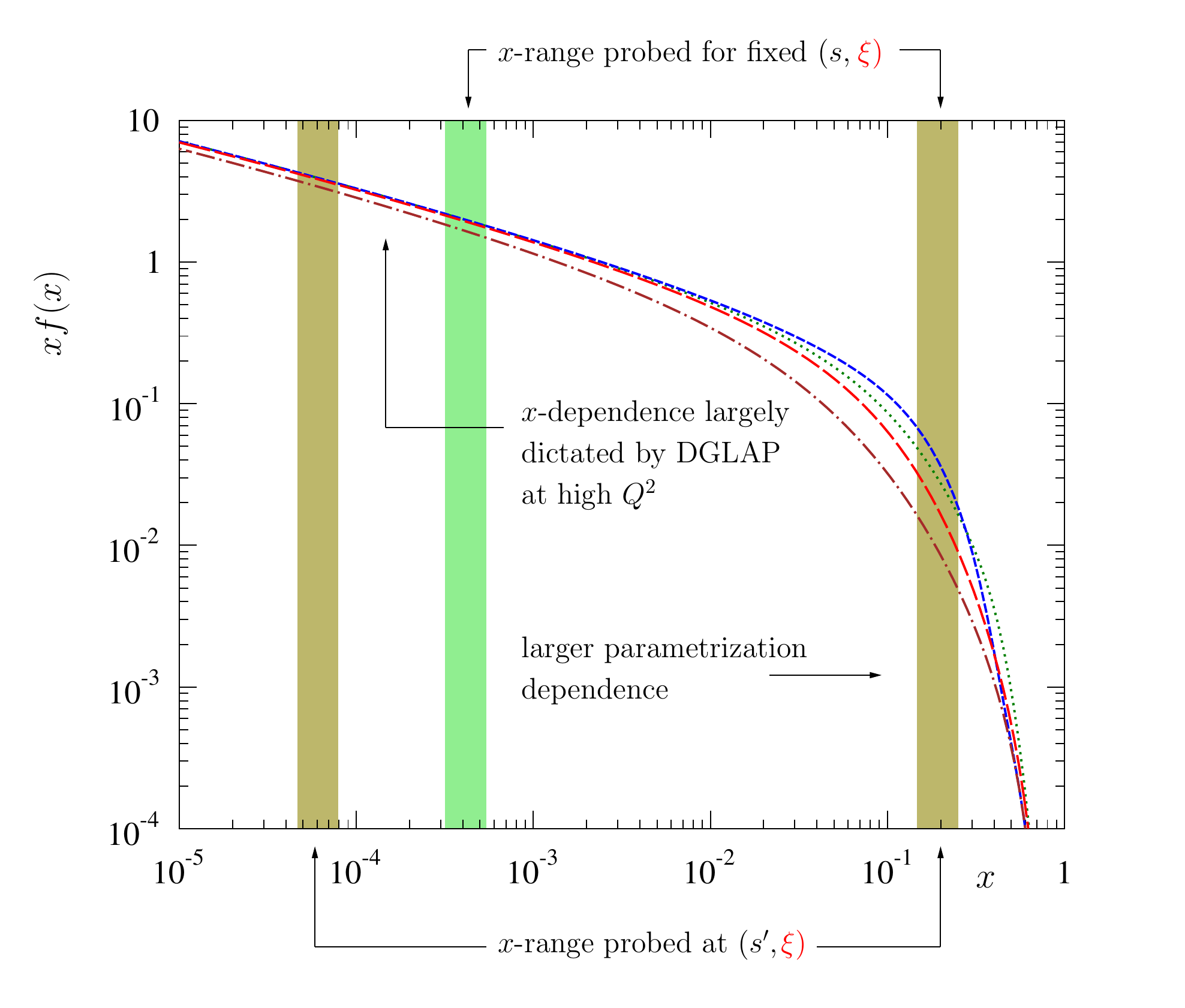}
\vspace{-0.2cm}\caption[]{A pictorial representation of the probed $x$ ranges with fixed rapidity $y$ (left) and fixed $\xi_1$ (right). The curves correspond to the CT10NLO PDFs \cite{Lai:2010vv} at $Q^2=M_{\rm W}^2$.}
\label{fig:Wscalingfordummies}
\end{figure}

In this talk, we summarize the main findings of our recent article \cite{Arleo:2015dba} in which we consider the inclusive production of W$^\pm$ bosons in the leptonic decay channel,
$$
{\rm H}_1 + {\rm H}_2 \rightarrow {\rm W}^- + {\rm X} \rightarrow \ell^- + \bar{\nu}
+ {\rm X},
$$
$$
{\rm H}_1 + {\rm H}_2 \rightarrow {\rm W}^+ + {\rm X} \rightarrow \ell^+ + \nu
+ {\rm X}.
$$
In particular, for a reason explained later, we will focus on the region with large leptonic rapidity $|y| \gg 0$. To understand what we are after, it is enough to examine the leading-order expressions for these processes in an approximation where the width $\Gamma_{\rm W}$ of W is much less than its mass $M_{\rm W}$. At particular center-of-mass (c.m.) energy $\sqrt{s}$, the cross sections differential in lepton rapidity $y$ and transverse momentum $p_{\rm T}$ read (see Ref.~\cite{Arleo:2015dba} for details),
{\small
{
\begin{eqnarray}
\frac{d^2\sigma^{\ell^\pm}(s)}{dydp_{\rm T}} & \approx & \frac{\pi^2}{24s} \left( \frac{\alpha_{\rm em}}{\sin^2\theta_{\rm W}} \right)^2 
\frac{1}{M_{\rm W} \Gamma_{\rm W}}
 \label{eq1}
 \frac{p_T}{\sqrt{1-4p_T^2/M_{\rm W}^2}} \sum _{i,j} |V_{ij}|^2 \, \delta_{e_{q_i} + e_{\overline{q}_j}, \pm1}   \\
& & \hspace{-2.8cm}
\left\{
\left[1 \mp \sqrt{1-4p_T^2/M_{\rm W}^2}\right]^2 q_i^{\rm H_1}({ x_1^+}) \overline{q}^{\rm H_2}_j({ x_2^+}) + \right. 
\left[1 \pm \sqrt{1-4p_T^2/M_{\rm W}^2}\right]^2 q_i^{\rm H_1}({ x_1^-}) \overline{q}^{\rm H_2}_j({ x_2^-}) + \nonumber \\
& & \hspace{-2.8cm}
\,\,\,\, \left[1 \pm \sqrt{1-4p_T^2/M_{\rm W}^2}\right]^2 \overline{q}^{\rm H_1}_j({x_1^+}) q_i^{\rm H_2}({x_2^+}) + 
\left.
\left[1 \mp \sqrt{1-4p_T^2/M_{\rm W}^2}\right]^2 \overline{q}^{\rm H_1}_j({x_1^-}) q_i^{\rm H_2}({x_2^-}) 
\right\}, \nonumber \\ \nonumber
\end{eqnarray}
}
}
where the momentum arguments of the PDFs $q_i^{\rm H_1}$ and $q_i^{\rm H_2}$ are
{\small
\begin{eqnarray}
 \hspace{-0.0cm}
 { x_1^\pm \equiv \frac{M_{\rm W}^2 e^y \,\,\,\, }{2p_T\sqrt{s}} \left[1 \mp \sqrt{1-4p_T^2/M_{\rm W}^2}\right]}, \quad \quad
 {x_2^\pm \equiv \frac{M_{\rm W}^2 e^{-y}}{2p_T\sqrt{s}} \left[1 \pm \sqrt{1-4p_T^2/M_{\rm W}^2}\right]}. \label{eq:x} 
 \end{eqnarray}
}
Symbols $\alpha_{\rm em}$, $\theta_{\rm W}$ and $V_{ij}$ denote the QED coupling, weak-mixing angle, and CKM matrix. The electric charges of the quarks are marked by $e_{q_i}$. Defining a scaling variable $\xi_1 \equiv \frac{M_{\rm W}}{\sqrt{s}}e^y$, we see that the momentum fractions $x_1^\pm$ can be written as
{\small
\begin{equation}
x_1^\pm \rightarrow \frac{M_{\rm W}}{2p_T} \xi_1 \left[1 \mp \sqrt{1-4p_T^2/M_{\rm W}^2}\right].
\end{equation}
}
This implies that the cross-sections at fixed $\xi_1$ are sensitive to the PDFs of hadron $H_1$ at particular values of $x$, independently of $\sqrt{s}$. The qualitative difference between considering cross sections at fixed rapidity or fixed $\xi_1$ is illustrated in Figure~\ref{fig:Wscalingfordummies}: If the rapidity $y$ is kept constant, the cross sections with two different c.m. energies $\sqrt{s}$ and $\sqrt{s'}$, probe the PDFs at different ranges of $x$ (left panel). However, if the scaling variable $\xi_1$ is maintained fixed instead, one samples the larger-$x$ PDFs (assuming $y>0$) at the same values of $x$ (right panel) irrespective of the c.m. energy.

By making some further simplifications, we are able to derive rather simple scaling laws. Approximating the small-$x$ PDFs at large interaction scale $Q^2$ by a power law
\begin{equation}
x\overline{q}_i(x,Q^2) \approx x{q}_i(x,Q^2) \approx N_i \ x^{-\beta(Q^2)}, \ x \ll 1, \label{eq:apr1}
\end{equation}
where $\beta \equiv \beta(Q^2=M_{\rm W}^2) \approx 0.35$ \cite{Arleo:2015dba}, it follows that
\begin{equation}
\frac{d\sigma^{\ell^\pm}(\sqrt{s},\xi_1)}{d\xi_1} \approx \left(\sqrt{s}\right)^{2\beta} \times F^\pm(\xi_1,{\rm H_1},{\rm H_2}), \quad y \gg 0, \label{eq:ab}
\end{equation}
where $F^\pm(\xi,{\rm H_1},{\rm H_2})$ is a function that does not depend explicitly on $\sqrt{s}$ or $y$. The approximation of Eq.~(\ref{eq:apr1}) should be reasonable if $x$ is small enough, which can be ensured by considering large $y$. 
We also note that at large $x$ the power-law approximation would not be a good one and this is the reason why, as in Figure~\ref{fig:Wscalingfordummies}, we intend to ``align'' the $x$ ranges at large-$x$ and not at small $x$. In the case of W charge asymmetry $\mathcal{A}(\xi_1,\sqrt{s},H_1,H_2)$, the $\sqrt{s}$ dependence cancels completely,
\begin{equation}
\mathcal{A}(\xi_1,\sqrt{s},H_1,H_2) = \frac{d\sigma^{\ell^+}/d\xi_1-d\sigma^{\ell^-}/d\xi_1}{d\sigma^{\ell^+}/d\xi_1+d\sigma^{\ell^-}/d\xi_1} \approx 
\frac{F^+(\xi_1,{\rm H_1},{\rm H_2})-F^-(\xi_1,{\rm H_1},{\rm H_2})}{F^+(\xi_1,{\rm H_1},{\rm H_2})+F^-(\xi_1,{\rm H_1},{\rm H_2})} \label{eq:asy1} \, ,
\end{equation}
even if the $\beta$ was $x$ dependent (as it definitely is). Making a further approximation that the behaviour of small-$x$ PDFs is flavor independent
\begin{equation}
x\overline{q}_i(x,Q^2) \approx x{q}_i(x,Q^2) \approx N \ x^{-\beta(Q^2)}, \quad x \ll 1, \label{eq:apr2}
\end{equation}
it follows that
\begin{equation}
\mathcal{A}(\xi_1,\sqrt{s},{\rm H}_1,{\rm H}_2) \approx F(\xi_1,{\rm H_1}), \qquad y \gg 0 \,. \label{eq:asy2}
\end{equation}
That is, at fixed value of $\xi_1$, the W charge asymmetry depends effectively only on the species (proton, nucleus, \ldots) probed at large $x$. As a consequence, the prediction is that on should be able to directly compare W charge asymmetry in e.g. p--p and p--Pb collisions at $y^{\ell} \gg 0$. The same scaling laws can obviously be derived also for $y<0$ using a scaling variable $\xi_2 \equiv \frac{M_{\rm W}}{\sqrt{s}}e^{-y}$.

\vspace{-0.4cm}
\section{Results}

\vspace{-0.3cm}
In what follows, instead of presenting plots as a function of $\xi_1$, we will use the shifted-rapidity variable $y_{\rm ref}$,
\begin{equation}
y_{\rm ref} \equiv y + \log \left( \frac{\sqrt{s_{\rm ref}}}{\sqrt{s}} \right), \nonumber
\end{equation}
where $\sqrt{s_{\rm ref}}$ is a chosen reference c.m. energy. For example, if we take $\sqrt{s_{\rm ref}}=7\,{\rm TeV}$, the rapidity variable at $\sqrt{s} = 8\,{\rm TeV}$ is shifted by 
$y \rightarrow y+\log(7{\rm TeV}/8{\rm TeV}) \approx y-0.134$. We note that a similar rapidity shift has been recently discussed also in the context of heavy-flavor production \cite{Gauld:2017omh}.

\begin{figure}[htb!]
\center
\includegraphics[width=0.35\textwidth]{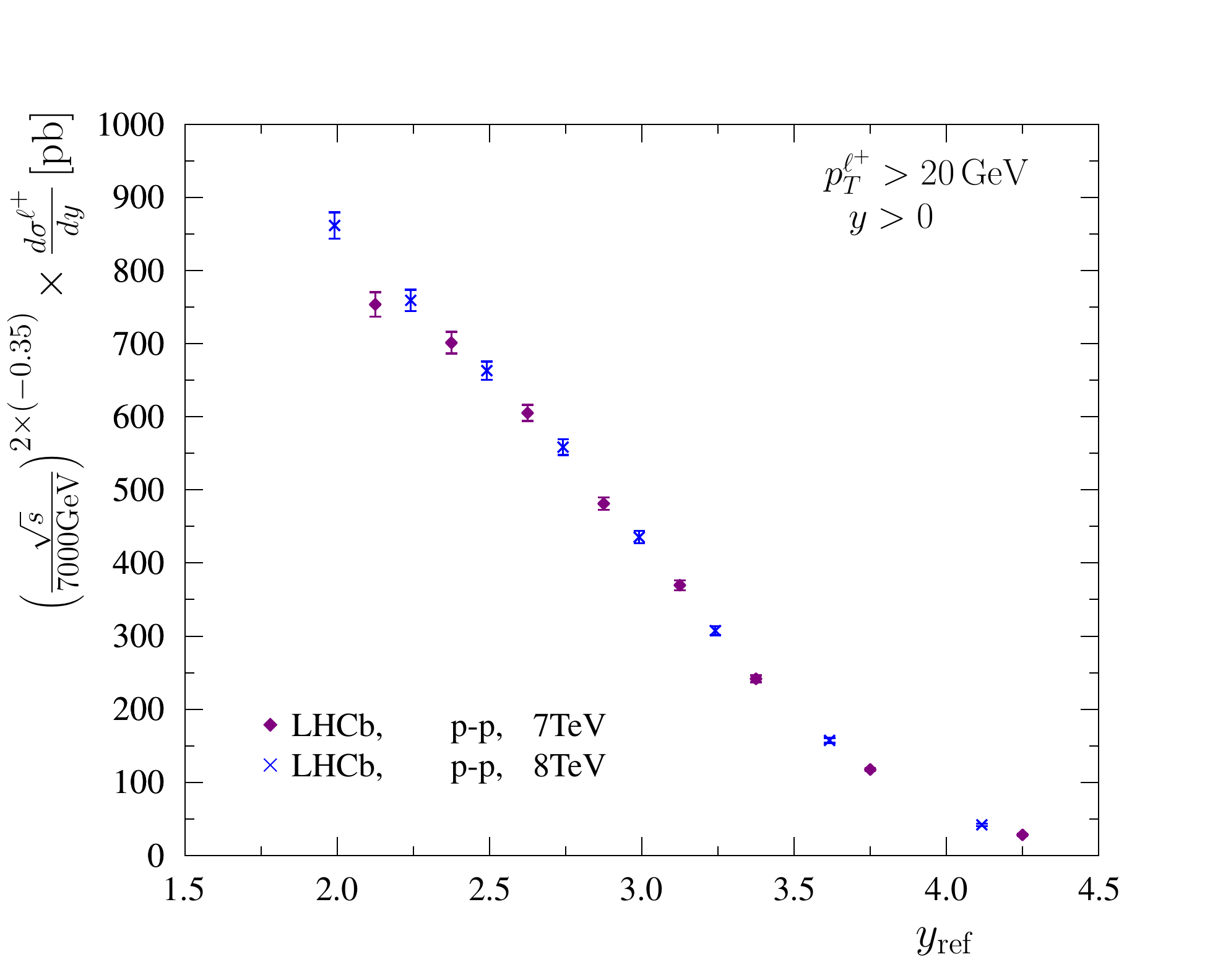}
\hspace{-0.6cm}
\includegraphics[width=0.35\textwidth]{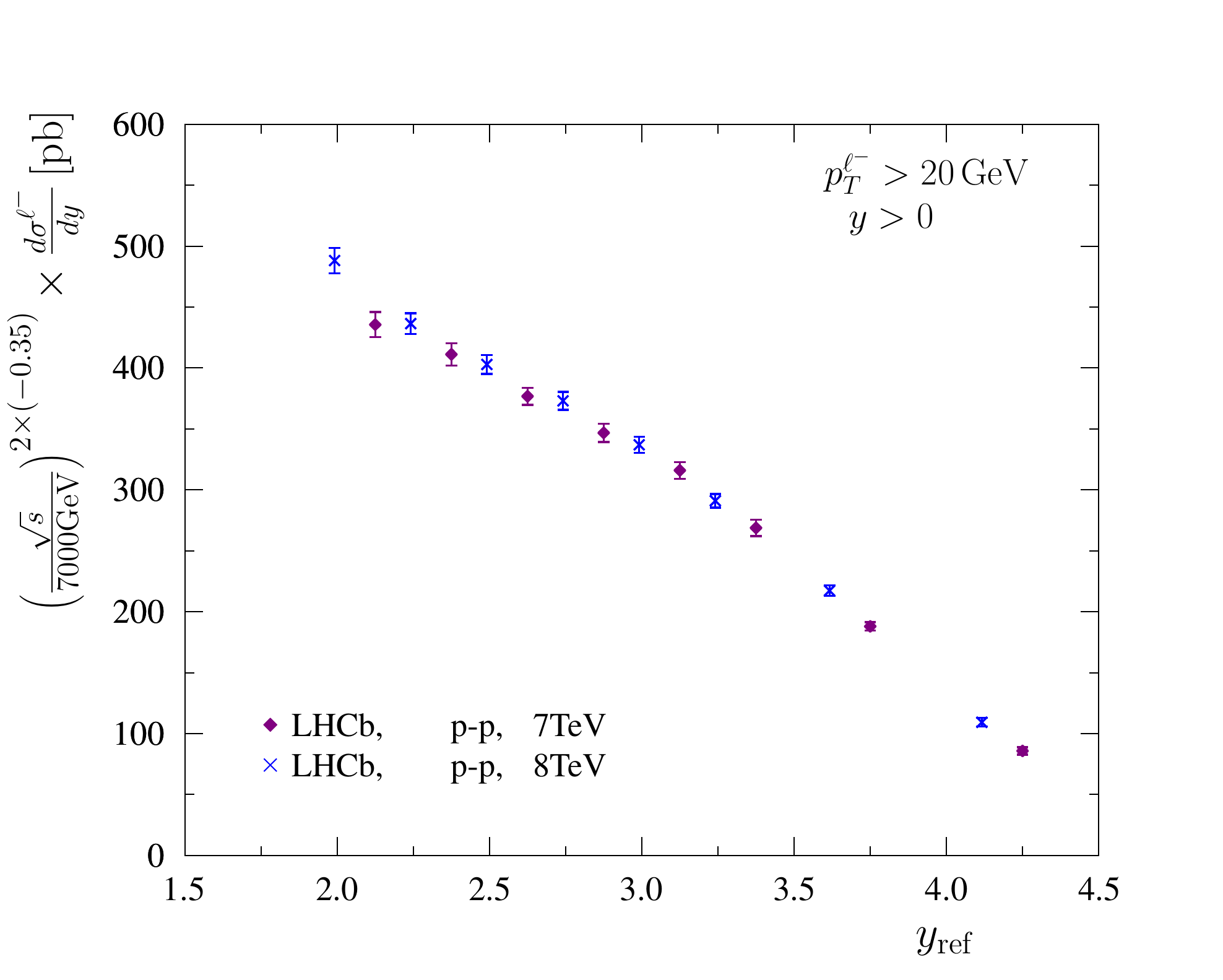}
\hspace{-0.6cm}
\includegraphics[width=0.35\textwidth]{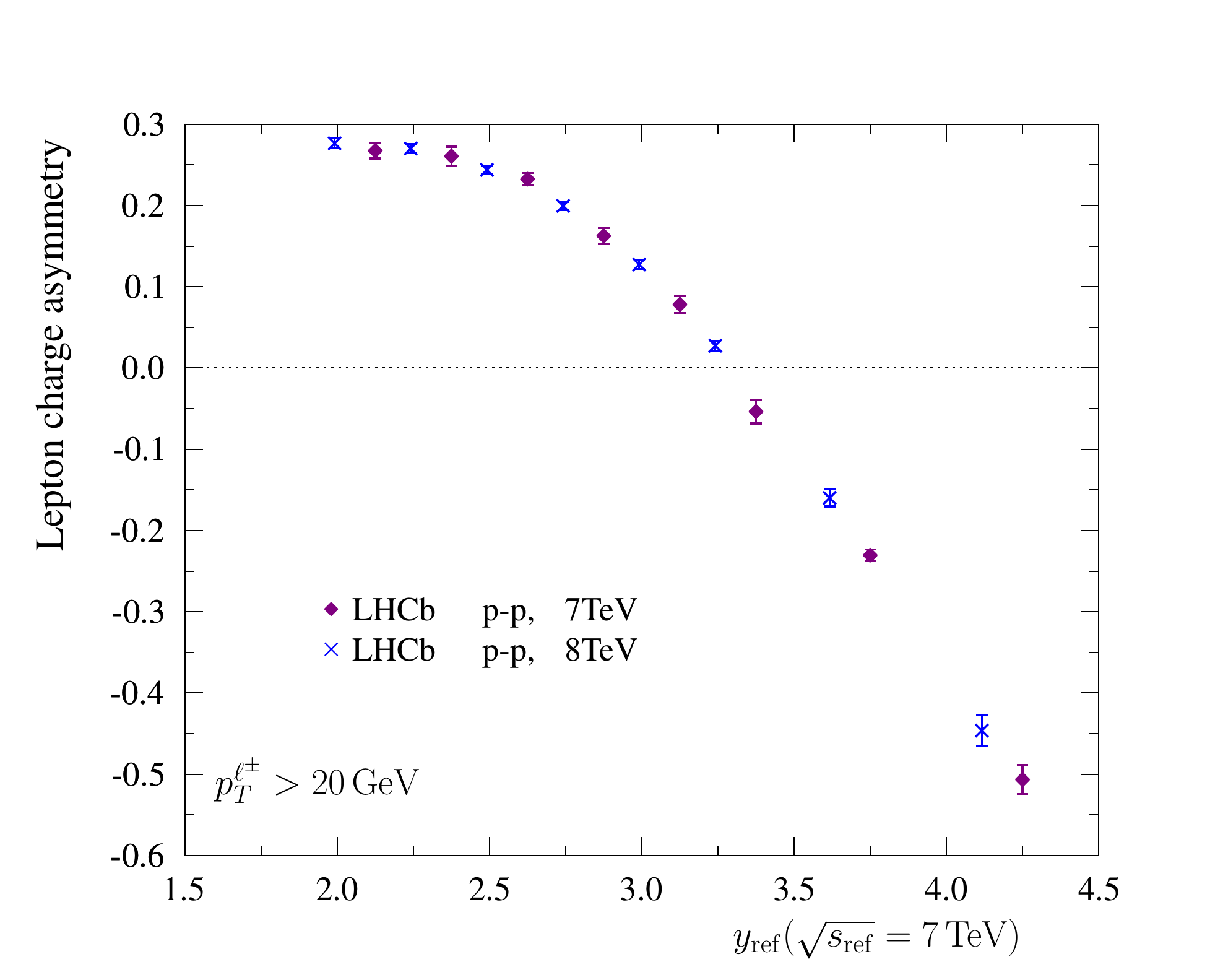}
\vspace{-0.2cm}\caption[]{The LHCb data at $\sqrt{s}=7,8\,{\rm TeV}$ and for W$^+$ (left), W$^-$ (middle), and charge asymmetry (right).}
\label{fig:LHCb}
\end{figure}

In Figure~\ref{fig:LHCb}, we contrast the LHCb $\sqrt{s}=7\,{\rm TeV}$ \cite{Aaij:2014wba} and $\sqrt{s}=8\,{\rm TeV}$ \cite{Aaij:2015zlq} p--p data against the derived scaling laws. In the case of absolute cross sections we plot the quantity $\left(\sqrt{s}/7{\rm TeV}\right)^{-2\beta} \times {d\sigma^{\ell^\pm}(\sqrt{s},y_{\rm ref})}/{dy_{\rm ref}}$, which, as far as Eq.~(\ref{eq:ab}) is accurate, should be independent of $\sqrt{s}$. As can bee seen from Figure~\ref{fig:LHCb} (left and middle panels), the data points indeed line up on a same curve to a very good approximation. The right-most panel of Figure~\ref{fig:LHCb} shows the W charge asymmetry as a function of $y_{\rm ref}$. In agreement with Eq.~(\ref{eq:asy1}), the data points settle roughly on a same curve. 

\begin{wrapfigure}{r}{0.64\textwidth}
\vspace{-1.2cm}
\includegraphics[width=0.64\textwidth]{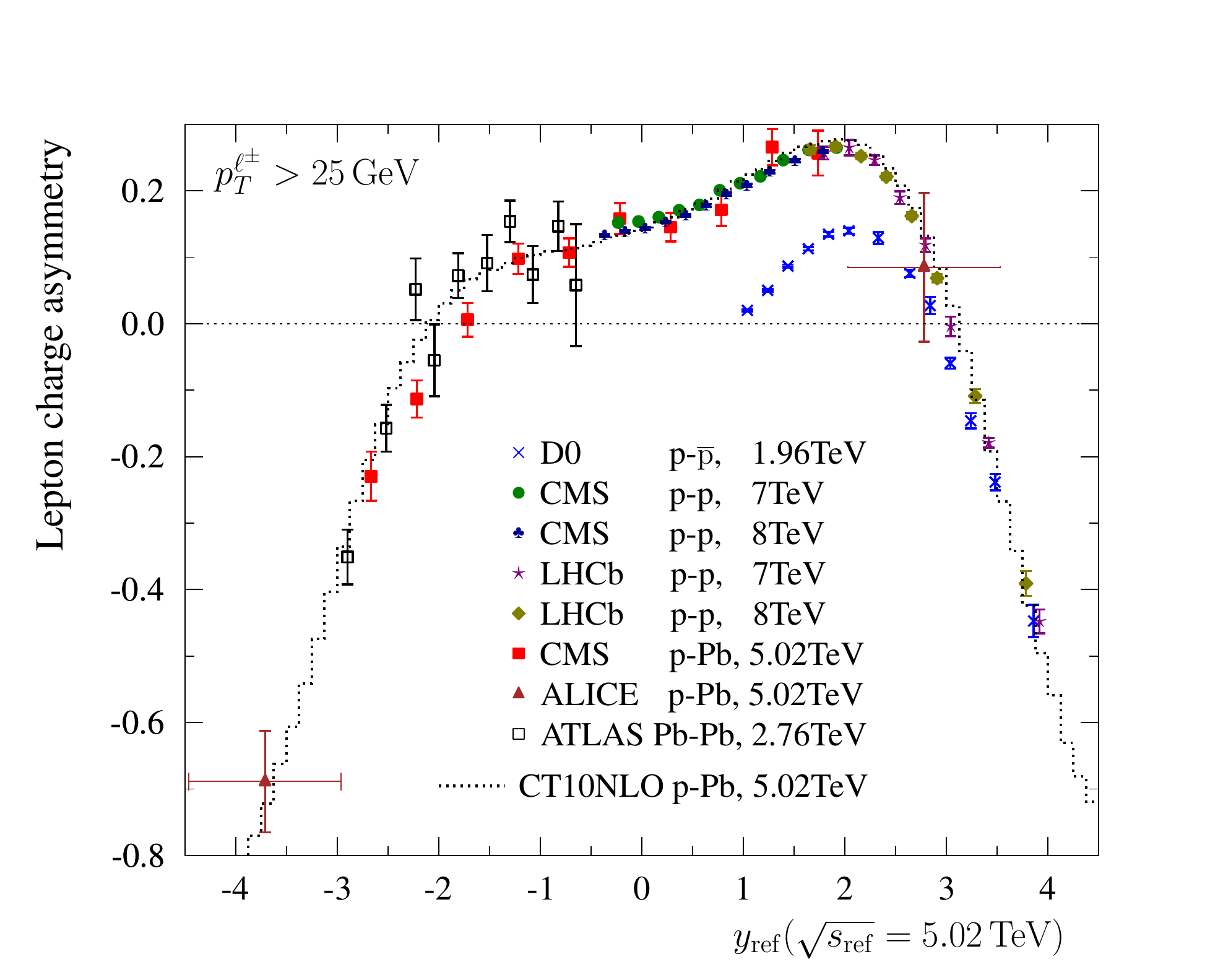}
\vspace{-0.8cm}\caption[]{The world data on charge asymmetry as a function of $y_{\rm ref}$.}
\label{fig:master}
\end{wrapfigure}

Finally, we test the prediction of Eq.~(\ref{eq:asy2}) against the world data on W charge asymmetry. This is done in Figure~\ref{fig:master} where we combine the data from p--p, p--$\overline{\rm p}$, p--Pb, and Pb--Pb collisions in a single plot. The prediction is that at fixed value of $y_{\rm ref}$,
\vspace{-0.0cm}
{\small
\begin{eqnarray}
y \gg 0 & : & \mathcal{A}({\color{red}\rm p}\overline{\rm p}) \approx \mathcal{A}({\rm {\color{red}p}p}) \approx \mathcal{A}({\rm {\color{red}p}Pb}) \ \ ({\rm probe \ {\color{red}p} \ at \ large \ }x) \nonumber \\
y \ll 0 & : & \mathcal{A}({\rm p{\color{blue} Pb}}) \approx \mathcal{A}({\rm Pb{\color{blue} Pb}}) \ \ \ \ \ \ \ \ \ \ \ \, ({\rm probe \ {\color{blue}Pb} \ at \ large \ }x) \nonumber
\end{eqnarray}
}
To keep the plot readable, the p--p and p--$\overline{\rm p}$ data are plotted only for $y>0$ and Pb--Pb data only $y<0$. Indeed, to a good approximation, the data align on curve which corresponds to the charge asymmetry in p--Pb collisions. In the case of Tevatron p--$\overline{\rm p}$ data one has to go to rather large $y_{\rm ref}$ to see the unification with the LHCb data as the $\sqrt{s}$ is lower and probed $x$ values too large for the approximation made in Eq.~(\ref{eq:apr1}) and Eq.~(\ref{eq:apr2}) to be accurate.

\begin{wrapfigure}{r}{0.45\textwidth}
\includegraphics[width=0.45\textwidth]{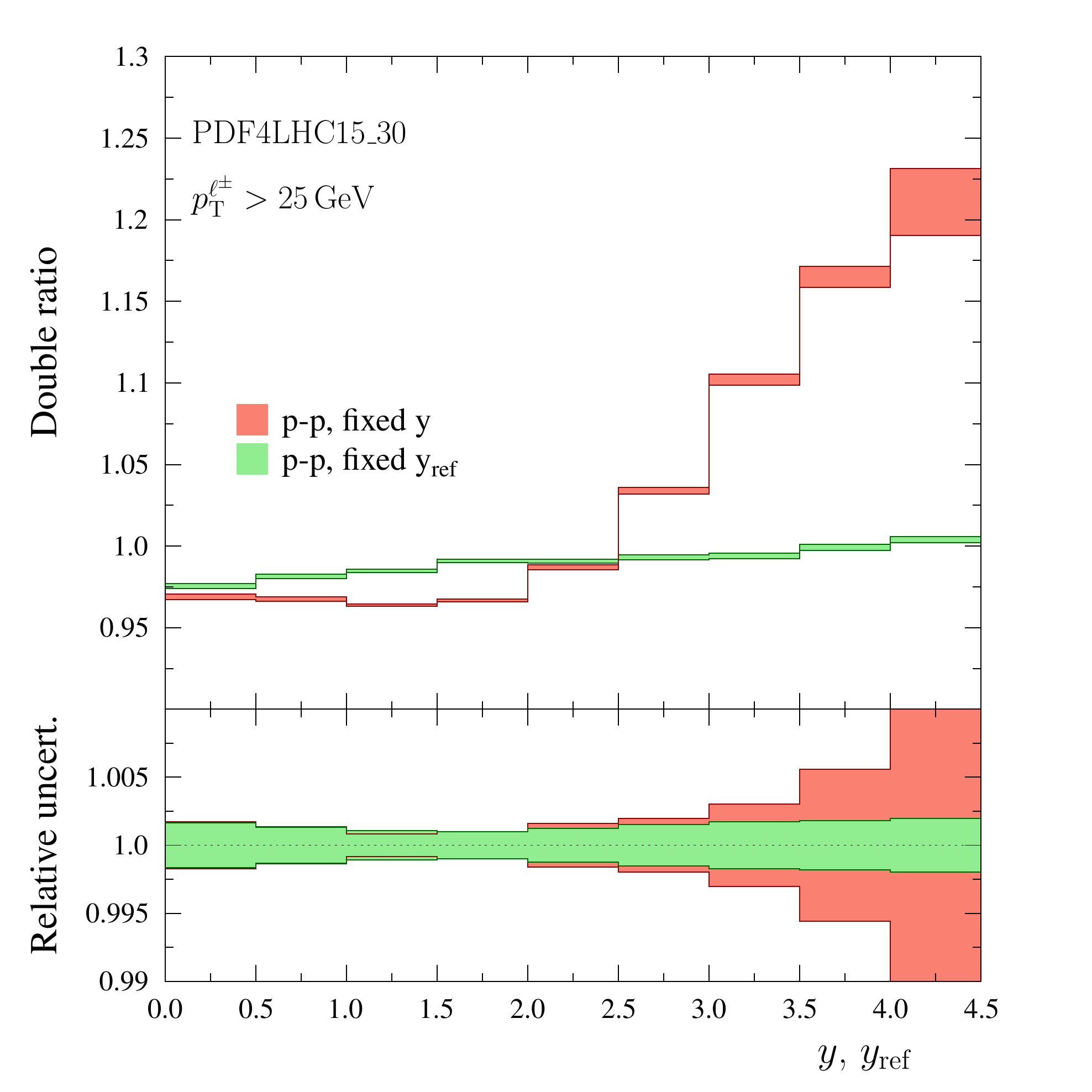}
\caption[]{The double ratio $\mathcal{D}_{8/7}$ at fixed rapidity $y$ (red) and at fixed $y_{\rm ref}$ (green). The bands represent the uncertainties from PDF4LHC15\_30 PDFs \cite{Butterworth:2015oua}.}
\label{fig:double}
\end{wrapfigure}

Having now seen that the derived scaling laws are indeed rather good estimations, we can turn them into a tool for precision physics. To this end, we consider the double ratio $\mathcal{D}_{8/7} = (R_{8/7}^+)/(R_{8/7}^-)$, where
\begin{equation}
R_{8/7}^+ = \frac{d\sigma^{\rm W^+}(\sqrt{s}=8\,{\rm TeV})}{d\sigma^{\rm W^+}(\sqrt{s}=7\,{\rm TeV})}, \quad
R_{8/7}^- = \frac{d\sigma^{\rm W^-}(\sqrt{s}=8\,{\rm TeV})}{d\sigma^{\rm W^-}(\sqrt{s}=7\,{\rm TeV})}.
\end{equation}
In Figure~\ref{fig:double}, we show the predictions as obtained by using PDF4LHC15\_30 set \cite{Butterworth:2015oua} of PDFs. As can be seen, the PDF uncertainty becomes clearly narrower when the ratio is formed at fixed $y_{\rm ref}$ than when the ratio is taken at fixed $y$. The smaller uncertainties follow from a better cancellation of the large-$x$ PDF uncertainties as now at fixed $y_{\rm ref}$ the probed large-$x$ regions are the same for both c.m. energies (see Figure~\ref{fig:Wscalingfordummies}). This observation should e.g. help in understanding the deviations from the NLO predictions for the integrated double ratio recently observed by the LHCb \cite{Aaij:2015zlq}.

\vspace{-0.4cm}
\section{Summary}

\vspace{-0.3cm}
We have reported on a study of scaling properties of inclusive W$^\pm$ production. The approximate scaling laws we have derived facilitate an easy understanding of the $\sqrt{s}$ dependence of the absolute cross sections and charge asymmetry. We have also demonstrated how one can directly compare measurements in different collision systems. As a by product, we have found that when the ratios of  W$^\pm$ cross sections between different $\sqrt{s}$ are taken at fixed value of the scaling variable $\xi_1 = \frac{M_{\rm W}}{\sqrt{s}}e^y$, the predictions are particularly robust against PDF errors and will facilitate tests of the Standard Model with reduced PDF uncertainty.

\vspace{-0.3cm}
\section*{Acknowledgments}

\vspace{-0.3cm}
\noindent The work of \'EC has been supported by the European Research Council, under the ``QuarkGluonPlasmaCMS'' \#259612 grant. H.P. acknowledges the funding from Academy of Finland, Project 297058; the European Research Council grant HotLHC ERC-2011-StG-279579 ; Ministerio de Ciencia e Innovaci\'on of Spain and FEDER, project FPA2014-58293-C2-1-P; Xunta de Galicia (Conselleria de Educacion) - H.P. is part of the Strategic Unit AGRUP2015/11.

\end{document}